# Drilling deep in South Pole Ice

Timo Karg [a)] and Rolf Nahnhauer [b)]

*Deutsches Elektronen Synchrotron, DESY*
*D-15738 Zeuthen, Platanenallee 6*

a) timo.karg@desy.de
b) Corresponding author: rolf.nahnhauer@desy.de

**Abstract.** To detect the tiny flux of ultra-high energy neutrinos from active galactic nuclei or from interactions of highest energy cosmic rays with the microwave background photons needs target masses of the order of several hundred cubic kilometers. Clear Antarctic ice has been discussed as a favorable material for hybrid detection of optical, radio and acoustic signals from ultra-high energy neutrino interactions. To apply these technologies at the adequate scale hundreds of holes have to be drilled in the ice down to depths of about 2500 m to deploy the corresponding sensors. To do this on a reasonable time scale is impossible with presently available tools. Remote drilling and deployment schemes have to be developed to make such a detector design reality. After a short discussion of the status of modern hot water drilling we present here a design of an autonomous melting probe, tested 50 years ago to reach a depth of about 1000 m in Greenland ice. A scenario how to build such a probe today with modern technologies is sketched. A first application of such probes could be the deployment of calibration equipment at any required position in the ice, to study its optical, radio and acoustic transmission properties.

## INTRODUCTION

Recently the IceCube Neutrino Observatory at the South Pole has observed the first cosmic neutrinos up to PeV energies opening this way a new window to study our universe at large distances [1].

IceCube consists out of 86 strings carrying 60 Digital Optical Modules deployed between 1450 m and 2450 m depth in the Antarctic Ice shield at the South Pole over a horizontal area of about one square kilometer. The successful construction of the detector was strongly dependent on the ability to drill large diameter holes with high speed down to the deep ice [2].

The IceCube drilling station finally allowed drilling a 2500 m deep hole of 60 cm diameter in less than 48 hours. A 5 MW power station was used to produce 880 l/min of hot (90°C) water at a pressure of 135 bar. With this hot-water-drill, 30 trained people were able to drill a maximum of 20 holes per season and to finish the deployment of 86 strings in seven years

The ARA experiment [3] under construction at the South Pole near to IceCube aims to detect even higher energetic cosmogenic neutrinos at much lower flux levels [4]. For ARA 37 clusters of 6 dry holes, each 16 cm wide and 200 m deep, should be drilled over a surface area of ≈ 100 km$^2$ to contain antennas for the detection of radio signals from neutrino interactions. The distance between clusters is 2000 m. For that purpose a transportable hot water drill has been constructed and has successfully been used in the 2012/2013 Antarctic season.[5]. Two dry holes could be drilled per day within two 12-hour shifts. But only 80 m of the holes are drilled in compact ice.

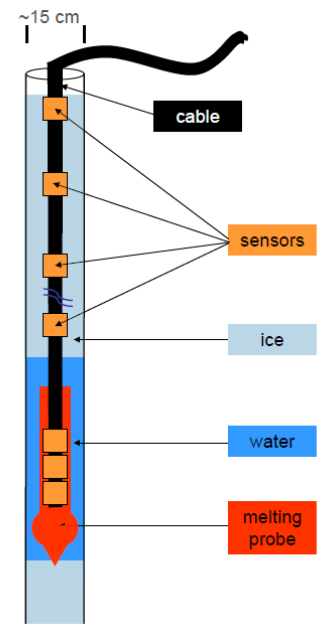

FIGURE 1: A schematic view of a possible string design and autonomous deployment

Experiments which use only one signal detection technology and sparse instrumentation have difficulties to separate the expected small signal from the large existing background. This is the reason to think about a hybrid detector using different signals from the same neutrino interaction to extract the wanted information with high significance. This leads to detector concepts with hundreds of sensor strings distributed at distances of 300 – 500 m down to depths of 1000 – 2500 m. The realization of such a kind of detector needs careful preparation with different ambitious R&D projects, e.g. for sensors, communication, power generation and distribution, but also for new drilling concepts. For several hundreds to thousands of strings going down to 1000 m depth or more probably only autonomous drilling will allow to realize such an experiment in a reasonable number of years. A possible string design with an autonomous melting probe at the bottom is schematically sketched in Fig. 1.

## THE PHILBERTH PROBE AND ITS SUCCESSORS

The German physicist Karl Philberth had already done the first successful steps in designing and testing autonomous thermal drilling probes in the 1960s [6,7]. He had designed probes consisting of tubes with 110 mm diameter and 2600 mm length. The metallic front part is used for electrical heating, melting the ice by about 50 m/day using 4 kW of power. The power is transmitted through wires, which are contained in the second part of the probe made of epoxy. The two probes tested in Greenland reached depths of 220 and 1000 m. The first probe stopped after 220 m due to a short cut between the wires for unknown reasons.

More detailed information about the construction of the probes, including a modified version using a different stabilization scheme for vertical drilling, is given in [8, 9] by H. W. C. Aamot. Details about the two probes which he describes in his paper are visible from Fig. 2. In contrast to Philberth, he mentions the use of power levels from 5 to 15 kW for penetration rates from 2.5 m up to 5 m per hour.

Until the mid-1990s several thermal probes have been built with small modifications to the original design. Tests have been done in Greenland and in Antarctica. A nice historical overview is given by J. Kelty in [10]. More recent reviews of the subject are given in [11] and [12].

During the 1990s autonomous melting probes have been designed and built also in Germany by the Alfred-Wegner Institute for Polar Research [13]. Little is known about corresponding field tests.

A basic modification in the design concept of thermal probes has been reported only in the year 2001 [14] in an IEEE paper from NASA engineers who have taken the task to develop a probe for use in space on planetary or moon surfaces. They suggested to combine direct thermal drilling with hot water drilling using the original melting water for that purpose.

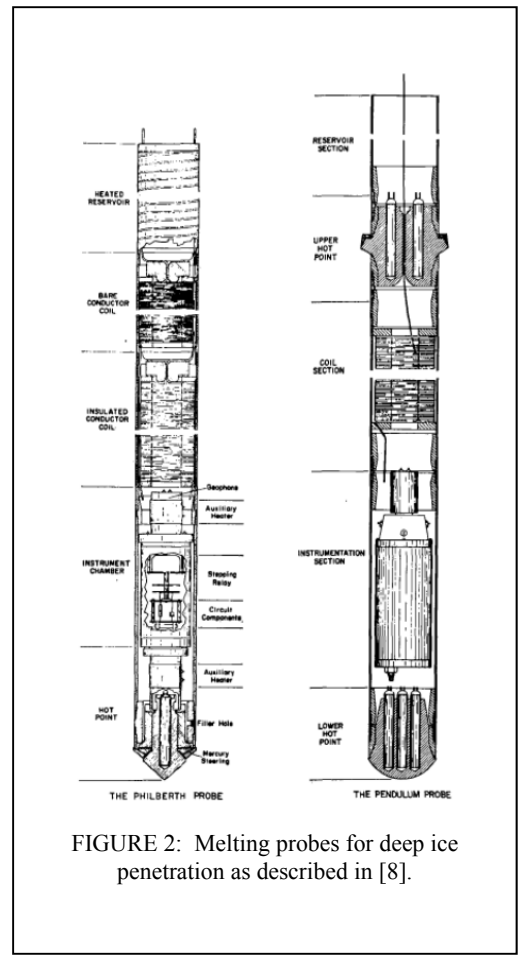

FIGURE 2: Melting probes for deep ice penetration as described in [8].

A recent concept, also for planetary applications, is described in the IceMole project [15] (now Enceladus Explorer project [16]). Using the IceMole basic design, first ideas have been discussed for deploying future sensors in Antarctic ice [17]. However, using an IceMole is giving much more possibilities than are necessary for a simple, vertical sensor deployment. A very different, ambitious approach is followed in the project VALKYRIE [18]. There, a high power laser is used to transmit energy for heating via a thin optical fiber to the melting head of a cryo-robot.

# BUILDING AN AUTONOMOUS MELTING PROBE TODAY

At DESY a group of people tried to design a melting probe [19] following Philberth's concept with up to date mechanical and electronic components. In this section we will give a short sketch of its basic parts.

The probe diameter is 160 mm and the overall length is presently 1420 mm. It consists of three sections, the melting head at the bottom, the sensor section which contains all steering electronic components as well as sensors for different applications, e.g. an acoustic positioning system and devices for studying ice properties, and finally a container for 3000 m of wire, used for power supply and communication between the probe and the surface. This last section has heater mats at the outside, to avoid freezing of the probe during movement in the ice. The wire is spooled to a coil using ortho-cyclic winding [20]. A special launch procedure has been developed to start melting in the firn layer at South Pole.

To calculate the power requirements for melting the probe down into the ice with a velocity v, we closely followed the considerations of Aamot in [21].

The total power used only for melting is:

$$P_m = \pi r^2 v \rho (H_m + c_p \Delta T),$$

It depends on the cross section of the probe of radius r, the melting speed v, the density of the ice $\rho$, the melting heat $H_m$, the specific heat $c_p$ and the temperature difference $\Delta T$. The lateral heat $P_l$ necessary to avoid that the wall of the probe freezes back to the ice is a function of the temperature, the melting speed, and the radius and length of the probe:

$$P_l = T v r^2 f(L/vr^2),$$

The values for $P_l$ for a certain speed v and our probe geometry are calculated using the parameterization given in [21. Aamot [21] suggests that the ratio $P_l/P_m$ should be smaller than 1. Choosing for melting down into the ice a speed of v = 2 m/h leads to a necessary melting power of ~5 kW. For a ratio of $P_l/P_m$ = 0.4, this leads to a lateral power requirement of 2 kW. The total necessary power is just the sum of both quantities. Therefore we get a total power requirement of the probe of $P_{tot}$ = 7 kW for a melting speed
of 2 m/h. Adding a contingency of about 15% for losses not taken into account here, we should supply a power of 8 kW to the probe.

To transfer 8 kW of power we use a power station at the surface with a fixed voltage of 2500 V DC. The wire is assumed to be a coaxial cable with an outer diameter of 2 mm. The inner conductor has a diameter of 1.08 mm, i.e. a cross section of 0.92 mm$^2$. It is surrounded by a solid dielectric of 0.3 mm thickness, carrying the non-insulated outer conductor with a wall thickness of 0.16 mm. The dimensions are chosen so that the inner and outer conductors have equal cross sections

The steering and data acquisition system of the probe consists of a steering and DAQ controller located inside the probe and a control CPU at the surface station. The two components communicate with each other using a digital protocol that is modulated onto the DC supply power line. The total data rate over the cable is less the 300 bytes/sec. A slow control system allows to monitor all important parameters of the probe. An emergency procedure avoids overheating.of all system parts. The surface computer allows manual and remote operation of the probe. It maintains a database where all sensor and status information of the probe are stored.

# DISCUSSION OF FIRST APPLICATIONS

Considering the above, it seems to be possible to build an autonomous melting probe at reasonable cost. A rough cost estimate shows that about 15000 Euro are required for the surface equipment and 18000 Euro are the price for building one probe.

With a speed of 2m/h the probe would need about 50-55 days to reach a depth of 2500 m, consuming 3250 l of fuel during this time. The person-power to deploy the probe is small. It could be launched by 2-3 persons and needs
further on only computer control. One surface system could be used to deploy several probes at once

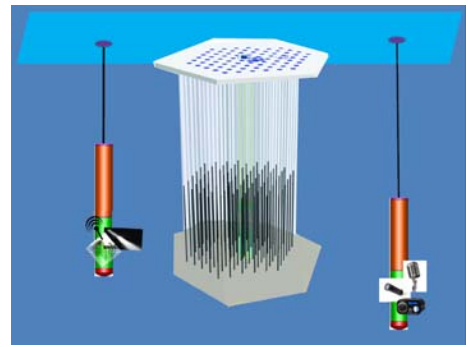

Figure 3: Schematic view of positioning calibration source emitters and receivers in the ice using melting probes

A direct comparison of the autonomous drilling concept with the IceCube drilling and deployment system is not possible here, because the problem of optical, radio or acoustic sensor deployment for neutrino detection using a melting probe has to be solved in a next step. The present device can however already be used to investigate ice properties for optical, radio and acoustic signal creation and propagation at any wanted position in the ice. A schematic view, how this may look like is given in Fig.3. The advantage of this technology is, that any position in the ice is accessible without big efforts which would be necessary, using the present IceCube drilling scheme [2].

## ACKNOWLEDGMENTS

The authors want to thank A. Donat, H. Lüdecke and F. Tonisch from DESY as well as T. Ullmann from Ferchau Engineering GmbH for common work particularly on the technical part of the Conceptual Design Report [19]. One of us (RN) thanks H. Örter for pointing him to the work of K. Philberth on melting probes.

## REFERENCES

1. **M.G. Aartsen et al., Science 342 (2013) 1242856**
2. **T. Benson et al., to appear in Ann. Glaciology V55, I68 „Ice Drilling Technology" , 2014**
3. **P. Allison et al., Astropart. Phys. 35 (2012) 457**
4. **K. Greisen, Phys. Rev. Lett. 16 (1966) 748, G. Zatsepin, F. Kusmin, JETP Lett. 4 (1966) 78**
   **V.S. Berezinsky, G.T. Zatsepin, Phys. Lett. B 28 (1969) 423**
5. **T. Benson, Polar Technology Conference 2013, Annapolis, http:// polarpower.org/PTC/list_2013.html**
6. **K. Philberth, Umschau in Wissenschaft und Technik 16 (1970) 516**
7. **K. Philberth, Proc. of Symposium on Ice-Core Drilling, Nebraska 1974, reproduced from "Ice-Core Drilling" edited by John F. Splettstoesser by permission of the University of Nebraska Press, 1976**
8. **H.W.C. Aamot, J. Glaciol. 7 (1968) 321**
9. **H.W.C. Aamot, Proc. of ISAGE, Hanover (NH), 1968, Int. Ass. of Scientific Hydrology Publication 86:63**
10. **J. Kelty, PhD Thesis, University of Nebraska, 1995**
11. **] S. Ulamec et al., Rev Environ Sci Biotechnol 6 (2007) 71**
12. **C. Bentley et al, in "Drilling in extreme environments", Wiley-VCH Verlag GmbH & Co. KGaA, (2009) 221**
13. **W. Jokat, H. Örter (eds.), Berichte zur Polarforschung, Alfred-Wegner-Institut, Bremerhaven (1997) 106**
14. **W. Zimmerman et al., IEEE Aerospace Conference 1 (2001) 311**
    **L. French et al., Proc. i-SAIRAS 2001, Canadian Space Agency, St-Hubert, Quebec, 2001**
15. **B. Dachwald et al., https://events.icecube.wisc.edu/contributionDisplay.py?sessionId=6&contribId=9&confId=34**
16. **D. Heinen, , to appear in Ann. Glaciology V55, I68 „Ice Drilling Technology" , 2014**
17. **K. Laihem, Nucl. Instrum. Meth. A 692 (2012) S192**
18. **W. Stone, http://www.stoneaerospace.com/**
19. **A. Donat et al., "A Thermal Probe for Melting Deep Holes in Antarctic Ice", Conceptual design report, DESY 2013, http:// http://www.ifh.de/tauros/IceDrill/IceDrill-CDR-Full-v5-woApp.pdf**
20. **H.W.C. Aamot, CRREL Special Report 128, 1969**
21. **H.W.C. Aamot, CRREL Technical Report 194, 1967**